\documentstyle[12pt,fleqn,epsf]{ioplppt} 

\begin{document}

\title{Black  Hole  Relics in String Gravity:
       Last Stages of Hawking Evaporation}

\author{S. Alexeyev${}^{1,2}$,
        A. Barrau${}^1$,
        G. Boudoul${}^1$,
        O. Khovanskaya${}^2$,
        M. Sazhin${}^2$}

\address{${}^1$ Institut des Sciences Nucleaires (CNRS/UJF),
                53 avenue des Martyrs,
                F-38026 Grenoble cedex, France}
\address{${}^2$ Sternberg Astronomical Institute (MSU),
                Universitetsky Prospekt, 13,
                Moscow 119992, Russia}

\begin{abstract}
The endpoint of black hole evaporation is a very intriguing problem of
modern   physics.   Based    on   Einstein-dilaton-Gauss-Bonnet   four
dimensional  string gravity  model  we show that  black  holes do  not
disappear  and should  become  relics at the  end  of the  evaporation
process. The possibility of experimental  detection  of  such  remnant
black holes is investigated. If they really exist, these objects could
be  a  considerable  part  of  the non  baryonic  dark  matter  in our
Universe.
\end{abstract}

%

\section{Introduction}

Theoretical physics faces nowadays a  great  challenge.  There is four
dimensional Standard Model  on one side (and the additional dimensions
are  not  required   to   explain  experimental  data)  together  with
inflationary cosmology  based  on additional scalar fields \cite{c01}.
On  the  other  side  there   is   the   completely   supersymmetrical
string/M-theory. Building links between those approaches \cite{c02} is
a very motivating goal of  modern  physics which could be achieved  by
the study of microscopic black holes.

As  General  Relativity  is  not renormalizable, its  direct  standard
quantization is  impossible.  To  build  a semiclassical gravitational
theory, the usual  Lagrangian should be generalized, which is possible
in different  ways. One of them is  to study  the action expansion  in
scalar  curvature,  {\it i.e.} higher order curvature corrections.  At
the level of second order, according to the perturbational approach of
string theory,  the most natural  choice is the 4D curvature invariant
Gauss-Bonnet term $S_{GB} =  R_{ijkl}R^{ijkl}  - 4 R_{ij}R^{ij} + R^2$
\cite{c03}.

With 4D action, it is not possible to consider only  $S_{GB}$ because,
being full derivative, it does not contribute to  the field equations.
It  must be  connected  it  with  a scalar  field $\phi$  to  make its
contribution dynamical. The following 4D effective  action with second
order curvature corrections can be built:
\begin{eqnarray*}
S & = & \int d^4 x \sqrt{-g} \Biggl[ - R + 2 \partial_\mu \phi
      \partial^\mu \phi
    + \lambda \xi(\phi) S_{GB} + \ldots \Biggr],
\end{eqnarray*}
where $\lambda$ is the string coupling constant. As  in cosmology, the
most simple generalization of  the  theory (a single additional scalar
field)  is  not  possible  because  while   dealing  with  spherically
symmetric solutions, the ``no-hair'' theorem restriction must be taken
into account.

Treating   $\phi$   as  a  dilatonic  field,  the  coupling   function
$\xi(\phi)$ is  fixed from the first  string principles and  should be
written $\exp(-2\phi)$ \cite{c04,c05}, which leads to :
\begin{eqnarray}\label{a11}
S & = & \int d^4 x \sqrt{-g} \Biggl[ - R + 2 \partial_\mu \phi
      \partial^\mu \phi
    + \lambda e^{-2\phi} S_{GB} + \ldots \Biggr].
\end{eqnarray}
Such type of actions  can be considered as one of the  possible middle
steps between General Relativity  and  Quantum Gravity. In this paper,
we show that this effective string gravity model and its solutions can
be applied for a description  of  the last stages of primordial  black
holes (PBH)  evaporation  \cite{c071,c072}  and suggests possible dark
matter candidates \cite{c08}. This should be understood in the general
framework of Gauss-Bonnet black hole (BH) theory \cite{c06,c07}.

The paper is  organized  as follows: In Section  II  we briefly recall
previously obtained results  and point out some new features important
for this study, Section III is devoted to the establishment of the new
Hawking  evaporation  law (especially for the detailed description  of
last stages of  Gauss-Bonnet BH evaporation),  in Section IV  we  show
that the direct  experimental registration of such PBHs is impossible,
Section V is  devoted  to  PBH relics as dark  matter  candidates  and
Section VI contains discussions and conclusions.

\section{Black hole minimal mass}

\subsection{Black hole minimal mass in pure EDGB model}

For the  sake of completeness, main  results from Ref.  \cite{c07} are
briefly repeated.

Starting from the action (\ref{a11}), a static, spherically symmetric,
asymptotically flat black hole solution is considered. One of the most
convenient choice of metric in this model is
\begin{eqnarray}\label{e21}
ds^2 = \Delta dt^2 - \frac{\sigma^2 }{\Delta } dr^2 - r^2
(d \theta^2 + \sin^2 \theta d \varphi^2),
\end{eqnarray}
where $\Delta=\Delta(r)$, $\sigma=\sigma(r)$.

Asymptotic expansion of the solution has the usual quasi-Schwarzschild
form,
\begin{eqnarray*}
\Delta (r \rightarrow \infty) & = & 1 - \frac{2M}{r}
+ O\biggl(\frac{1}{r}\biggr), \\
\sigma (r \rightarrow \infty) & = & 1 - \frac{1}{2} \frac{D^2}{r^2}
+ O\biggl(\frac{1}{r^2}\biggr), \\
\phi   (r \rightarrow \infty) & = & \frac{D}{r}
+ O\biggl(\frac{1}{r}\biggr),
\end{eqnarray*}
where $M$ and $D$  are  ADM (Arnowit-Dieser-Misner) mass and dilatonic
charge respectively.  Using a dedicated code,  a BH type  solution was
obtained.   This    solution    provides    a   regular   horizon   of
quasi-Schwarzschild  type  and  the  asymptotic  behaviour  near  this
horizon $r_h$ is:
\begin{eqnarray}\label{a12}
\Delta & = & d_1 (r-r_h) + d_2 (r-r_h)^2 + \ldots, \nonumber \\
\sigma & = & s_0 +s_1 (r-r_h) + \ldots, \nonumber \\
\phi & = & \phi_{00} + \phi_1 (r-r_h) + \phi_2 (r-r_h)^2  + \ldots,
\end{eqnarray}
where $(r-r_h) \ll 1$, $s_0$, $\phi_0 = e^{-2\phi_{00}}$ and $r_h$ are
free independent parameters.

After  solving  the  equations  to the first perturbation  order,  the
following limit on the minimal BH size can be obtained:
\begin{eqnarray}
r_h^{inf} = \sqrt{\lambda} \ \sqrt{4 \sqrt{6}}  \phi_h (\phi_\infty),
\end{eqnarray}
where  $\lambda$  is a combination of the  string  coupling  constants
({\it fundamental  value})  and  $\phi_h  (\phi_\infty)$  is dilatonic
value  at  $r_h$,  depending  upon dilatonic value at  infinity  which
cannot be determined only in the framework of this model. According to
this formula and taking into account the numerical values, the minimal
BH  mass  has the order of  Planck  one (more precisely $\approx  1.8$
Planck masses \cite{c07}).

It is necessary to point out  that the stability of the solution under
time perturbations at the event  horizon  was  described in \cite{c09}
and was studied at the singularity $r_s$ in \cite{c10}.

Contribution of  higher  order  curvature  corrections  was studied in
\cite{c07,c122}  to  show  that,  in the bosonic case  with  heterotic
string models  (the question is still open  in SUSY  II), all the  new
topological  configurations   are   located   inside  the  determinant
singularity  and,  therefore, do not produce any  new  {\it  physical}
consequences. Our conclusions  remain valid when the next higher order
curvature corrections are made of pure products of Riemannian tensors.
This topic is under additional investigation now.

Gathering these results,  it  can be  concluded  that the solution  is
stable in all the particular points, and, therefore, at all the values
of initial data set.

\subsection{Effects of moduli fields}

Generalizing   the   model  by  taking  into  account  the   effective
contribution of additional  compact dimensions in the most simple form
--- scalar field --- the action must be given as
\begin{eqnarray}\label{231}
S & = & \int d^4 x \sqrt{-g} \Biggl[ - R + 2 \partial_\mu \phi
      \partial^\mu \phi
    + 2 \partial_\mu \psi \partial^\mu \psi
    + \Biggl(\lambda_\phi e^{-2\phi} + \lambda_\psi \xi(\psi)\Biggr)
      S_{GB} \nonumber \\
  & + & \mbox{ higher order curvature corrections} \Biggr].
\end{eqnarray}

This model was  studied into the  details in \cite{c11}.  For  current
investigations it should be  emphasized  that when the contribution of
moduli field  value is considered, a  naked singularity can  appear if
the size of  additional  dimensions is greater than  the  BH size. The
minimal BH mass must therefore  be  increased to 10 Planck masses  (to
avoid being in naked singularity region). It is a key  feature because
it  allows  to  move  away from  the Planck  region  and  to  use {\it
semiclassical   approach}.   If   these  additional  dimensions   were
non-compact \cite{c12} the BH minimal mass would be much greater.

\section{Black hole evaporation law}

\subsection{Probability of transition to the last stage}

According to the analysis given  in  Ref.  \cite{c130}, the transition
from  prelast  to  last  stage  of  BH evaporation  is  forbidden  and
evaporating PBHs will never reach the minimal mass state. The shape of
the BH  mass loss rate law changes  and becomes  the one presented  in
Fig.2,  analogously  to the  simplified  ``toy  model''  presented  in
\cite{c130}. Different types of similar models for BH evaporation were
studied in  Lovelock  gravity  \cite{c131},  string inspired curvature
expansions  \cite{c132}  and  in  many other theories.  The  numerical
values of Gauss-Bonnet BH (important for experimental search analysis)
will be presented in section D.

\subsection{Approximation to metric functions}

In  the  WKB   approximation   of  the  Hawking  evaporation  process,
everything happens  in the neighbourhood  of the event horizon. As our
metric functions $\Delta$  and  $\sigma$ depend upon radial coordinate
$r$ and black hole mass $M$, i.e. $\Delta = \Delta(M,r)$ and $\sigma =
\sigma  (M,r)$  (other  variables  are  not  important),  we  can  use
expansions  (\ref{a12}),  taking  into  account only the  first  terms
(partially neglecting  the  dependence  upon  radial  coordinate $r$).
Using \ref{a12}, the metric can be written as:
\begin{eqnarray}\label{a31}
\Delta (M,r) & = & 1 - \frac{2 M }{r} \epsilon (M)
               = \frac{1}{2 M \epsilon} (r - 2 M \epsilon (M)),
                                                          \nonumber \\
\sigma (M,r) & = & \sigma_0 (M).
\end{eqnarray}

Using  the  numerically  calculated  data,  fits  were  performed  for
$\epsilon (M)$ and $\sigma_0 (M)$. As we are interested mostly  in the
last  stages  of  PBHs  evaporation,  where  the  difference  from the
standard Bekenstein-Hawking picture is considerable, Taylor expansions
around $M_{min}$  can be used. This also helps  in obtaining good fits
of the metric functions (see Fig.3 and Fig.4) which can  be considered
as  polynom  expansions (of  $M$  or  $1/M$)  that  are  valid between
$M=M_{min}=10\ M_{Pl}$ and $M=1000 \ M_{Pl}$ with good accuracy.
\begin{eqnarray}\label{e33}
\epsilon & = & 1 - \frac{\epsilon_1}{M} - \frac{\epsilon_2}{M^2}
               + \frac{\epsilon_3}{M^3} - \frac{\epsilon_4}{M^4}, \\
\sigma_0 & = & \sigma_2 (M-M_{min})^2
       - \sigma_3 (M-M_{min})^3
       + \sigma_4 (M-M_{min})^4
       - \sigma_5 (M-M_{min})^5, \nonumber
\end{eqnarray}
where (for $M_{min} = 10 \ M_{Pl}$) the corresponding coefficients are
$\epsilon_1 = 10.004$,  $\epsilon_2  = 13.924$, $\epsilon_3 = 2856.3$,
$\epsilon_4 = 25375.0$, $\sigma_2  =  0.11933 * 10^{-04}$, $\sigma_3 =
0.30873  *  10^{-07}$, $\sigma_4 = 0.30871 *  10^{-10}$,  $\sigma_5  =
0.11051 * 10^{-13}$.

Using this  technique, the PBH  evaporation spectra and mass loss rate
were derived in  an  analytical form  (valid  only near the  $M_{min}$
point).

\subsection{Black hole evaporation spectra in EDGB model}

In some  approaches, black holes are treated as  immersed in a thermal
bath and the evaporation can  be  described as a WKB approximation  of
semiclassical   tunnelling   in   a   dynamical   geometry.   In   our
investigation, we follow  the  techniques described in \cite{c139} and
\cite{c14}.  The  same method was also applied  in  \cite{c14a}.  Some
other descriptions of BH evaporation can be found in \cite{c15,c17}.

The key idea of the method from Ref.\cite{c139} and \cite{c14} is that
the  energy of a  particle  changes  its  sign when  crossing  the  BH
horizon. So, a pair created  just  inside or just outside the  horizon
can become  real with  zero total energy after one  member of the pair
has tunnelled  to the opposite side.  The energy conservation  plays a
fundamental  role:  transitions  between  states with the  same  total
energy are the only possible ones. Using quantum  mechanical rules, it
is possible to write the imaginary  part of the action for an outgoing
positive  energy  particle  which  crosses the horizon  outwards  from
$r_{in}$ to $r_{out}$ as:
\begin{eqnarray*}
Im (S)   =   Im   \int_{M}^{M    -    \omega}   \int_{r_{in}}^{r_{out}}
\frac{dr}{\dot r} dH,
\end{eqnarray*}
where $\omega$ is the energy of the particle, $H$ is total Hamiltonian
(and  total energy)  and the  metric  is written  so as  to avoid  the
horizon  coordinate  singularity.   Following  \cite{c14},  Painleve's
coordinates  are  used.  The  transformation to this metric  from  the
Schwarzschild one can be obtained by changing the time variable:
\begin{eqnarray*}
t = t_{old} + r\sqrt{ \frac{\sigma^2}{\Delta^2}-\frac{1}{\Delta} }.
\end{eqnarray*}
Substituting $t_{old}$ into (\ref{e21}) one obtains
\begin{eqnarray}\label{e31}
ds^2 & = & - \Delta dt^2 + 2  \sqrt{\sigma^2 -  \Delta} dr  dt +  dr^2
+ r^2 d \Omega^2.
\end{eqnarray}

In WKB approximation, the  imaginary  part of the semiclassical action
$Im(S)$, describing the probability of tunnelling  through the horizon
is
\begin{eqnarray}\label{e40}
Im(S) = Im   \int\limits_{r_{in}}^{r_{out}}   p_r   \,dr   =   Im
\int\limits_{r_{in}}^{r_{out}} \int\limits_0^{p_r} p_r' \,dr,
\end{eqnarray}
where $p_r$ is canonical momentum.

For Gauss-Bonnet BH the radial geodesics are described by the equation
\cite{c07}
\begin{eqnarray}\label{e41}
\dot{r} =  \frac{dr}{d\tau}  = \frac{\Delta}{ \sqrt{\sigma^2 - \Delta}
\mp \sigma} = \mp \sigma - \sqrt{\sigma^2 - \Delta}.
\end{eqnarray}

After substituting  expression (\ref{e41}) to the equation (\ref{e40})
one obtains:
\begin{eqnarray}\label{e44}
Im (S) & =  & Im     \int\limits_M^{M     -     \omega}
\int\limits_{r_{in}}^{r_{out}} \frac{dr}{\dot r} dH
 = - Im \int\limits_0^\omega \int\limits_{2M  \epsilon}^{2  (M  -
\omega)\epsilon}  \frac{dr  d\omega'}{  \sigma   -   \sqrt{\sigma^2  -
\Delta}}.
\end{eqnarray}

Substituting the expression (\ref{a31}) extended in (\ref{e33}) to the
equation (\ref{e44}), the imaginary part of the action  can be written
as:
\begin{eqnarray*}
Im(S)  &  = & - Im  \int\limits_{0}^{\omega} d\omega'
\Biggl( \int\limits_{2M
\epsilon}^{2 (M - \omega')\epsilon}
\frac{dr}{   \sigma  -   \sqrt{\sigma^2   -   \frac{r}{2   (M   -
\omega')\epsilon} + 1}} \Biggr)
\end{eqnarray*}
Changing variables with
\begin{eqnarray*}
y = \sqrt{ \sigma^2 - \frac{r}{2(M-\omega)\epsilon} + 1 }.
\end{eqnarray*}
$Im(S)$ takes the form
\begin{eqnarray*}
Im(S) & =  & - Im    \int\limits_{0}^{\omega}     d\omega'    \Biggl(
\int\limits_{\sqrt{\sigma^2 - \frac{\omega'}{M  -  \omega'}}}^{\sigma}
\frac{4 (M - \omega')\epsilon y dy}{y - \sigma} \Biggr) \\
&   =  &  -Im \int\limits_{0}^{\omega} d\omega'  \Biggl( 4 (M  -  \omega')
\epsilon\sigma \int\limits_{\sqrt{\sigma^2   -   \frac{\omega'}{M   -
\omega'}}}^{\sigma} \frac{dy}{y - \sigma} \Biggr) \\
&  = &  -  \int\limits_{0}^{\omega} d\omega' \left(  4  (M -  \omega')
\epsilon \sigma \pi \right).
\end{eqnarray*}

As a result the imaginary part of the action is :
\begin{eqnarray*}
2 Im(S) = \frac{840 \pi}{M^2(M-\omega)^2} \alpha,
\end{eqnarray*}
where $\alpha$ is a huge  expression  that cannot be written here.  It
can be found at http://isnwww.in2p3.fr/ams/ImS.ps.

Using the  numerical values for  a realistic order of $M_{min}$ around
10 Planck  masses, the corresponding $\epsilon_{i},$ and $\sigma_{j}$,
it is possible to find from (\ref{a31}) the approximate expression for
$Im(S)$.  As  we  are  interested  mostly in  the  last  stages  of BH
evaporation where the influence of higher  order curvature corrections
is  important,  the  limit  $M-M_{min}  \ll 1$  can  be  taken  in the
computations, leading to a  very  different spectrum than the standard
Bekenstein-Hawking  picture  (where $ - dM/dt \propto 1/M^2$).  Taking
into account energy  conservation, $\omega$ can  be bounded :  $0  \le
\omega  \le  M  -  M_{min}$.  The  approximate  expression  of  $Im(S)
(M,\omega)$ for a given $M_{min}$ can then be used in the form
\begin{eqnarray}\label{e49}
Im(S) =  k * (M-M_{min})^3,
\end{eqnarray}
where  constant  $k=5\cdot10^{-4}$  in  Planck  unit   values  with  a
satisfying accuracy (the plot of $Im(S)$  and  its  approximation  are
shown on Fig.5).

\subsection{Energy conservation and mass lost rate}

Following Ref.\cite{c14}, the emission spectrum per  degree of freedom
can simply be written as:
\begin{eqnarray}
\frac{d^2N}{dEdt}=\frac{\Gamma_s}{2\pi\hbar}\cdot
\frac{\Theta((M-M_{min})c^2-E)}{e^{Im(S)}-(-1)^{2s}},
\end{eqnarray}
$\Gamma_s(M,E)$ being the absorption probability  for  a  particle  of
spin $s$ and  the Heavyside function  being implemented to  take  into
account energy conservation with a minimal  mass  $M_{min}$.  In  this
section and in the following ones, standard units are used  instead of
Planck ones as numerical  results  should be obtained for experimental
fluxes. At this point, two questions have to be addressed:  which kind
of fields are  emitted (and which  correlative $\Gamma_s$ have  to  be
used) and which mass range is physically interesting.  To answer those
questions, the mass loss rate is needed:
\begin{eqnarray}
-\frac{dM}{dt}=\int_0^{(M-M_{min})c^2}\frac{d^2N}{dEdt}\cdot
\frac{E}{c^2} dE
\end{eqnarray}
where the  integration is carried out up to  $(M-M_{min})c^2$ so as to
ensure  that  the   transition   below  $M_{min}$  is  forbidden.  The
absorption probabilities can  clearly be taken is the limit $GME/\hbar
c^3 \ll 1$ as we are considering the endpoint emission when the cutoff
imposed by $M_{min}$ prevents the  black  hole  from emitting particle
with  energies  of  the  order  of  $kT$.  Using  analytical  formulae
\cite{c19} and expanding $\exp(Im(S))$ to the  first  order  with  the
approximation according  to (\ref{e49}), it is  easy to show  that the
emission of spin-1 particles, given by (per degree of freedom)
\begin{eqnarray}
-\frac{dM}{dt}\approx\frac{16}{9\pi}\frac{G^4M_{Pl}}{\hbar^5c^2k}M^4
(M-M_{min})^3
\end{eqnarray}
dominates over s=1/2 and  s=2  emission whatever the considered energy
in the previously quoted limit.  It  is interesting to point out  that
the fermion emission around $M_{min}$ is not strongly  modified by the
EDGB model as, in the lowest order, $\exp(Im(S))-(-1)^{2s} \approx 2$.
Furthermore, if energy conservation was implemented as a simple cutoff
in the Hawking spectrum, the opposite result would  be obtained: s=1/2
particles would dominate the mass loss  rate as the power of $(ME)$ in
the  absorption  probability  is  the  smallest  one.  If  we restrict
ourselves to massless particles, {\it i.e.} the only ones emitted when
$M$ is  very close to $M_{min}$,  the metric modification  changes the
endpoint emission nature from neutrinos to photons. The real mass loss
rate  is  just  twice  the   one   given  here  to  account  for   the
electromagnetic helicity states.

With this expression $-dM/dt=f(M)$, it is possible to compute the mass
$M$ at any given time $t$ after formation at mass $M_{init}$ as:
\begin{eqnarray}
t=\int_{M}^{M_{init}}\frac{dM}{f(M)}\approx\frac{9\pi k\hbar^5
c^2}{32G^4M_{Pl}^3}
\times\frac{1}{M_{min}^4(M-M_{min})^2}
\end{eqnarray}
where only the  dominant term in  the limit $t\rightarrow  \infty$  is
taken from the analytical primitive of the function.  As expected, the
result does not depend on $M_{init}$ which is due to the fact that the
time needed  to go from $M_{init}$ to  a few  times $M_{min}$ is  much
less than  the  time  taken to go from a few times $M_{min}$ to $M$ as
long as $M_{init}<<10^{15}~{\rm g}$ for $t\approx 10^{17}~{\rm s}$. At
time $t$ after formation, the mass is given by:
\begin{eqnarray}
M\approx M_{min}+\sqrt{\frac{9k\pi\hbar^5 c^2}
{8M_{min}^4G^4M_{Pl}^3t}}
\end{eqnarray}

This mass can be implemented in the emission spectrum formula:
\begin{eqnarray}
\frac{d^2N}{dEdt}\approx\frac{32}{3\pi}\left(\frac{8}{9\pi}\right)^{
\frac{3}{2}}
G^{10}\hbar^{-\frac{25}{2}}c^{-15}M_{Pl}^{\frac{15}{2}}M_{min}^{10}
 \nonumber \\
k^{-\frac{5}{2}}\times t^{\frac{3}{2}} E^4\Theta\left(
\sqrt{\frac{9k\pi\hbar^5 c^6}{8M_{min}^4G^4M_{Pl}^3t}}-E\right)
\end{eqnarray}
leading to a frequency f given by
\begin{eqnarray}
f=\int_0^{(M-M_{min})c^2}\frac{d^2N}{dEdt}dE\approx\frac{36}{15}
\cdot\frac{1}{t}.
\end{eqnarray}

If we want  to investigate the  possible relic emission  produced  now
from PBHs  formed in the early universe with  small masses, this leads
to a frequency around $6\cdot 10^{-18}~{\rm Hz}$ with a typical energy
of  the  order of $1.8\cdot 10^{-6}$  eV.  This emission rate is  very
small  as  it   corresponds  to  the  evaporation  into  photons  with
wavelength much  bigger that the radius of the  black hole. It should,
nevertheless, be  emphasized  that  the  spectrum  is a monophonically
increasing  function  of  energy,  up  to  the cutoff,  with  a  $E^4$
behaviour.  Furthermore,  it  shows  that,  although   very  small  in
intensity, the  evaporation never stops  and leads to a mass evolution
in $1/\sqrt{t}$.

\section{Experimental detection}

We  investigate  in  this  section  the  possibility  to  measure  the
previously given relic  emission.  Let $R$  be  the distance from  the
observer, $z$ the redshift corresponding to the distance $R$, $\theta$
the opening  angle of the detector  (chosen so that  the corresponding
solid  angle  is  $\Omega=1~sr$),   $d^2N/dEdt(E,t)$   the  individual
differential  spectrum  of a  black  hole  relic  (BHR)  at  time $t$,
$\rho(R)$ the numerical BHR density taking  into  account  the  cosmic
scale  factor  variations,  $R_{max}$  the horizon in  the  considered
energy range, $t_{univ}$ the age  of  the Universe and $H$ the  Hubble
parameter. The "experimental" spectrum $F$ (J$^{-1}\cdot $s$^{-1}\cdot
$sr$^{-1}$) can be written as:
\begin{eqnarray}
F=\int_0^{Rmax}\frac{d^2N}{dEdt}\left(E(1+z),t_{univ}-\frac{R}{c}\right)
\nonumber \\
\times \frac{\rho(R) \cdot \pi R^2 tan^2(\theta) }{4\pi R^2} {\rm d}R
\end{eqnarray}
which leads to:
\begin{eqnarray}
F= \cdot tan^2(\theta) \frac{8}{3\pi}\left(\frac{8}{9\pi}
\right)^{\frac{3}{2}} G^{10}\hbar^{-\frac{25}{2}}c^{-15}
M_{Pl}^{\frac{15}{2}}M_{min}^{10}
\nonumber \\
\times k^{-\frac{5}{2}}  E^4\int_0^{R_{max}}\rho(R)
\left(\frac{1+\frac{HR}{c}}{1-\frac{HR}{c}}\right)^2
(t_{univ}-\frac{R}{c})^{\frac{3}{2}}
\nonumber \\
\times \Theta\left(\sqrt{\frac{9k\pi\hbar^5c^6}{8M_{min}^4G^4M_{Pl}^3
(t-\frac{R}{c})}}
-E\sqrt{\frac{1+\frac{HR}{c}}{1-\frac{HR}{c}}}\right){\rm d}R
\nonumber \\
\end{eqnarray}

This integral can be analytically computed and takes into account both
the  facts  that  BHRs far away from Earth must be taken at an earlier
stage of  their evolution and that  energies must be  redshifted. Even
assuming     the     highest     possible     density     of      BHRs
($\Omega_{BHR}=\Omega_{CDM}\approx  0.3$)  and  $R_{max}$  around  the
Universe   radius,    the   resulting   flux   is   extremely   small:
$F\approx1.1\cdot 10^{7}~{\rm J}^{-1}~{\rm  s}^{-1}~{\rm  m}^{-2}~{\rm
sr}^{-1}$ around $10^{-6}$ eV, nearly 20 orders of magnitude below the
background. This closes  the  question about possible direct detection
of BHRs emission.

Another  way  to investigate  differences  between  EDGB  black  holes
particle emission and  a  pure Hawking spectrum is  to  study the mass
region where $dM/dt$  is maximal. Taking  into account that  the  mass
loss rate becomes  much higher  in the  EDGB  case than  in the  usual
Hawking picture,  it could have  been expected that the extremely high
energy flux was strongly enhanced. In particular, it  could revive the
interest in PBHs as candidates to solve the enigma of  measured cosmic
rays above the GZK cutoff.  Nevertheless,  the  spectrum  modification
becomes  important only  when  the mass is  quite  near to  $M_{min}$.
Depending  on  the real  numerical  value of  $M_{min}$,  it can  vary
substantially (increasing with increasing $M_{min}$) but remains a few
Planck masses above $M_{min}$. This is  far too small to account for a
sizeable increase  of the flux.  The number of particles emitted above
$10^{20}$eV in a  pure  Hawking  model is of the  order  of  $10^{15}$
\cite{c072}. This value should be taken  with care as it relies on the
use of leading log QCD computations  of  fragmentation  functions  far
beyond the energies reached by colliders but the order of magnitude is
correct. On the other hand, even if all the energy available when EDGB
modifications  becomes   important   were  released  in  $10^{20}$  eV
particles (which is not realistic) it would generate only a  few times
$10^9$ particles and modify by less than 0.01\% the pure Hawking flux.
It would not allow to  generate,  as expected, a spectrum harder  than
$E^{-3}$.

\section{Primordial black holes as dark matter candidates}

The idea of PBH relics as a serious candidate for cold dark matter was
first  mentioned  in \cite{c08}.  It  was  shown  that  in  a Friedman
universe  without  inflation,  Planck-mass  remnants  of   evaporating
primordial black holes could be expected to have close to the critical
density.  Nevertheless,  the  study  was based on  the  undemonstrated
assumption  that  either a  stable  object forms  with  a mass  around
$M_{Pl}$ or a naked space-time singularity is left. Our study provides
new arguments favouring massive  relic  objects, probably one order of
magnitude above Planck  mass  and could  revive  the interest in  such
non-baryonic dark matter  candidates. An important problem is still to
be addressed in standard inflationary cosmology: the rather large size
of  the  horizon at  the  end  of  inflation.  The  standard formation
mechanism  of  PBHs  requires the mass of the black holes to be of the
order of the horizon mass at the formation time and only those created
after inflation  should be taken into account as  the huge increase of
the scale  factor would extremely  dilute all the ones possibly formed
before. It is easy to show  that under such assumptions the density of
Planck relics is very small:
\begin{eqnarray}
\Omega_{Pl}=\Omega_{PBH}\frac{\alpha-2}{\alpha-1}M_H^{1-\alpha}
M_*^{\alpha-2}M_{min}
\end{eqnarray}
where  $\Omega_{PBH}$  is  the  density  of  PBHs not yet  evaporated,
$\alpha$ is  the spectral index of the initial  mass spectrum (=5/2 in
the standard  model for a  radiation dominated Universe), $M_*$ is the
initial  mass of  a  PBH  which  evaporating time  is the  age  of the
Universe ($\approx 5\cdot 10 ^{14}$g) and $M_H$ is the horizon mass at
the end of inflation. This latter can be expressed as
\begin{eqnarray}
M_H=\gamma^{\frac{1}{2}}\frac{1}{8}\frac{M_{Pl}}{t_{Pl}}t_i
\approx \gamma^{\frac{1}{2}}\frac{1}{8}\frac{M_{Pl}}{t_{Pl}}
\frac{0.24}{(T_{RH}/1{\rm Mev})^2}
\end{eqnarray}
where  $T_i$  is the  formation  time and  $T_{RH}$  is the  reheating
temperature. Even with the highest possible value for $T_{RH}$, around
$10^{12}$  GeV  (according  to  Ref.  \cite{c20a}   if  the  reheating
temperature is more than $10^{9}$ GeV, BHR remnants  should be present
nowadays)   and   the  upper  limit  on  $\Omega_{PBH}$  coming   from
gamma-rays, around $6\cdot 10^{-9}$ the resulting density is extremely
small: $\Omega_{Pl}\approx 10^{-16}$.

There are, nevertheless,  at least two  different ways to  revive  the
interest in PBH dark  matter. The first one is related to  relics that
would  be  produced  from  an  initial  mass spectrum decreasing  fast
enough, so  as to overcome the gamma-rays limit.  The second one would
be  to  have  a  large  amount of  big  PBHs,  between  $10^{15}$g and
$10^{25}$g, where experiments  are  completely blind: such black holes
are too heavy to undergo Hawking  evaporation and too light to be seen
by microlensing experiments  (mostly because of the finite size effect
\cite{c21}).  The  most  natural  way  to  produce  spectra  with such
features is inflationary  models with a scale, either corresponding to
a change in  the  spectral index  of  the fluctuations power  spectrum
\cite{c22} either corresponding to a step \cite{c23}.

\section{Discussion and conclusions}

In this paper,  the  BH type solution of  4D  effective string gravity
action  with  higher order curvature corrections was  applied  to  the
description of  BHRs. A corrected  version of the evaporation law near
the minimal BH mass was  established.  It was shown that the  standard
Bekenstein-Hawking  evaporation   forluma  must  be  modified  in  the
neighbourhood of the last stages. Our main conclusion is to  show that
contrarily to what is usually thought the evaporation does not  end up
by the emission of  a few quanta with energy around Planck  values but
goes  asymptotically  to  zero  with an infinite  characteristic  time
scale.

The direct experimental registration of the products of evaporation of
BHRs is  impossible. This gives  an opportunity to consider these BHRs
as one of the main candidates for cold dark matter in our Universe.

\section*{Acknowledgments}

S.A. would  like to  thank the AMS Group in  the Institut des Sciences
Nucleaires (CNRS/UJF) de Grenoble for  kind  hospitality.  A.B. \& G.B
are very grateful to the Sternberg Astronomical Institute for inviting
them. This work  was supported in  part by ``Universities  of  Russia:
Fundamental Investigations'' via grant No. UR.02.01.026 and by Russian
Federation  State   Contract  No.  40.022.1.1.1106.  The  authors  are
grateful  to  A. Starobinsky  and  M. Pomazanov  for  the very  useful
discussions on the subject of this paper.


\begin{figure}
\epsfxsize=0.35\hsize
\centerline{\epsfbox{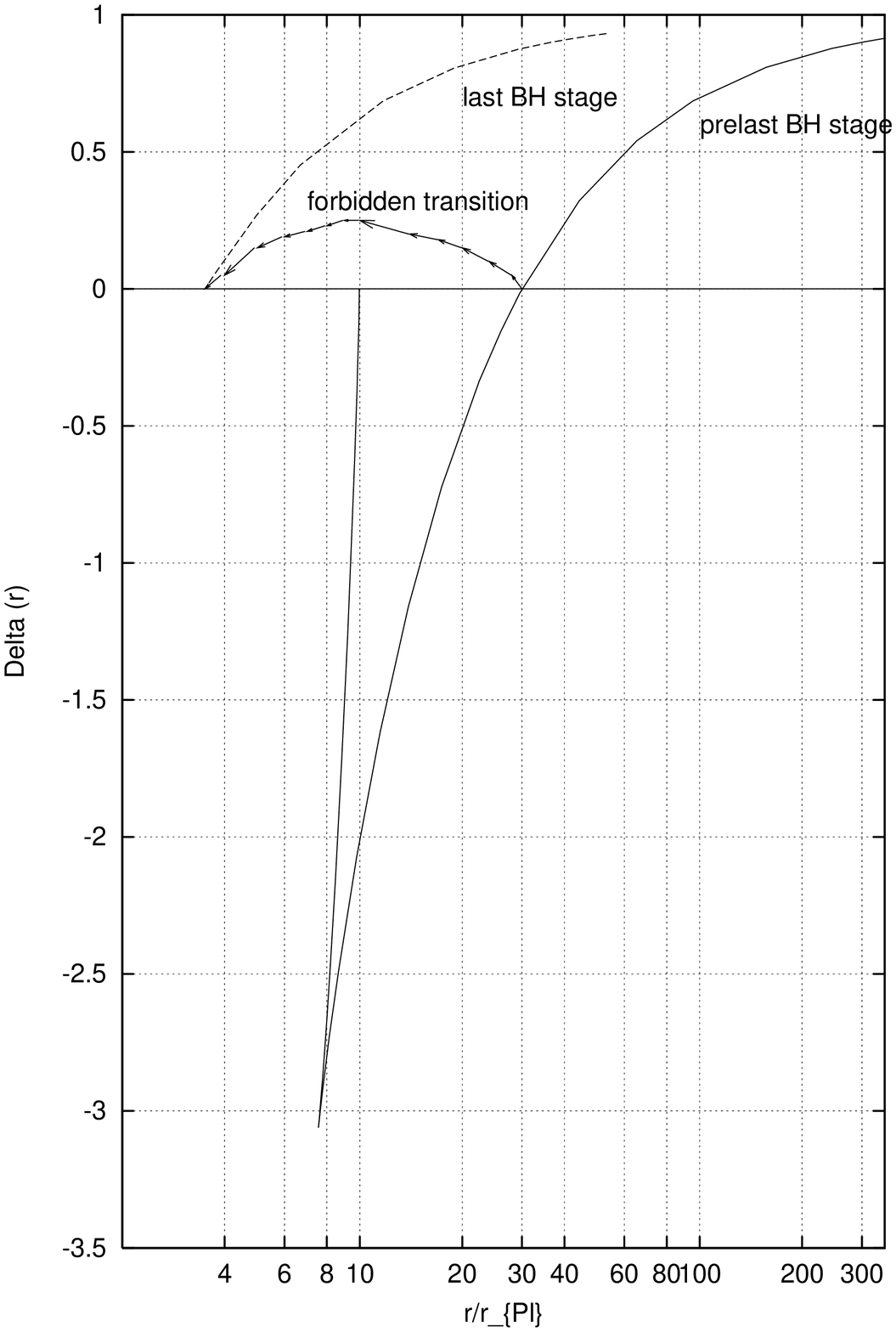}}
\caption{Illustration of the moment of last  transition. Prelast state
is   characterized   by   the   regular   horizon   with   the   usual
quasi-Schwarzschild  configuration.   The   last   state  is  singular
configuration  making  the  transition  from  prelast  to  last  state
forbidden.}
\end{figure}

\begin{figure}
\epsfxsize=0.35\hsize
\centerline{\epsfbox{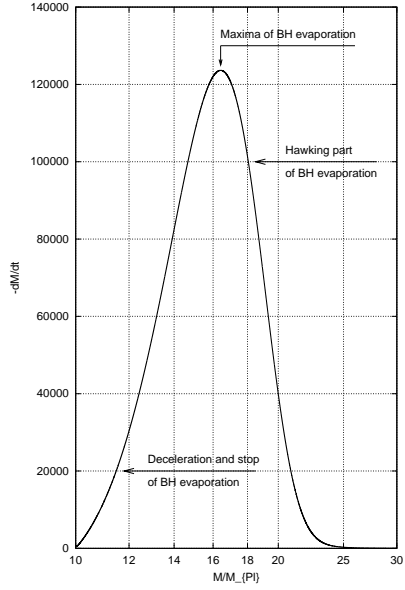}}
\caption{Shape of BH  mass lost rate  versus BH mass  in  Gauss-Bonnet
case when the energy conservation is taken into account. Right part of
the graph represents the usual  Hawking  evaporation  law when $-dM/dt
\sim  1/M^2$.  Left  part  shows  the  picture  at  last  stages  when
evaporation decelerates  and  then  stops,  distinguishing the minimal
possible mass (``ground state'').}
\end{figure}

\begin{figure}
\epsfxsize=0.35\hsize
\centerline{\epsfbox{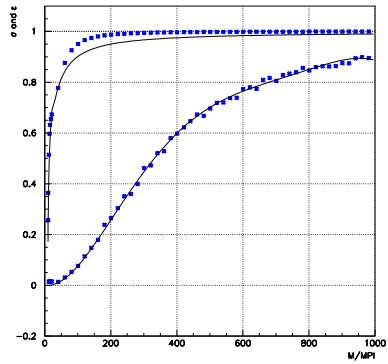}}
\caption{Metric function $\sigma$  and $\epsilon$ as a function of the
mass $M$ in Planck units for a fixed minimal mass $M_{Min}=10 M_{Pl}$.
Stars are numerically computed values and  the line is the fit used to
derive the spectrum.}
\end{figure}

\begin{figure}
\epsfxsize=0.35\hsize
\centerline{\epsfbox{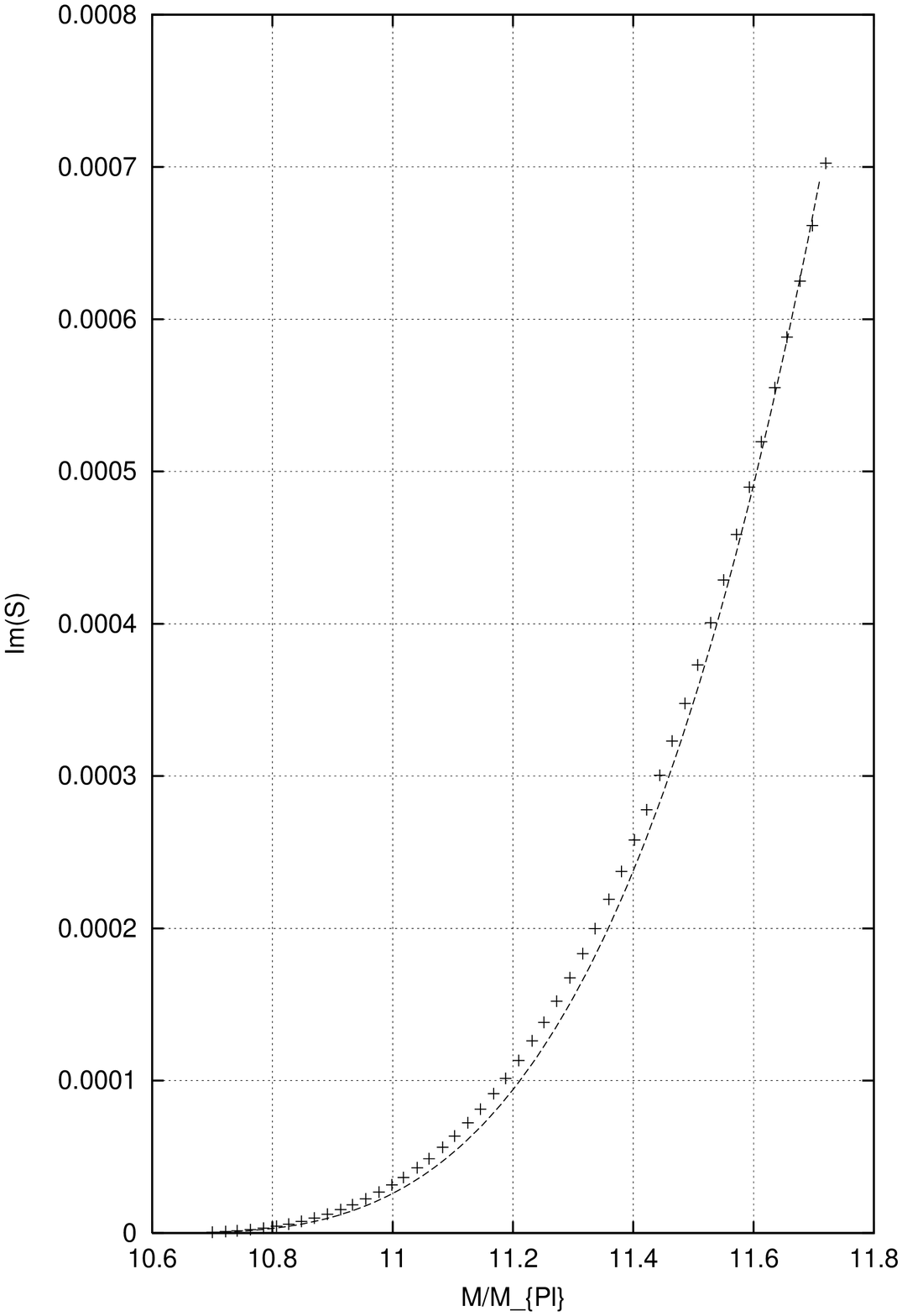}}
\caption{  $Im(S)$  (dots)  and  the  fit   $(5   \cdot   10^{-4})   *
(M-M_{min})^3$ (dashed line) versus BH mass $M$ during the last stages
of  BH  evaporation in  the  Gauss-Bonnet case  with  $M_{min} =  10.6
M_{Pl}$.  It  is  necessary  to  note  that  during   last  stages  of
evaporation the  emitted energy $\omega  < M-M_{min} \ll 1$. For fixed
values  of  $\omega   =  \omega^*_i$  in  the  vicinity  of  $M_{min}$
($O(M_{min})=0.01$)   the   mass   $M   \in   (    M_{min}+\omega^*_i,
M_{min}+\omega^*_i+O(M_{min})  )$.   So,   for   different  values  of
$\omega^*$ ($\omega^*_{i+1}  =  \omega^*_i  + O(M_{min}), \omega^*_1 =
0.1,  i  \in  N$)  $M$  belongs  to  different  (without intersection)
intervals.  Finally,  $Im(S)$  is  represented as connection  of  such
intervals with  the most probable  values of $\omega^*_i \in (0.1, 1.1
)$. }
\end{figure}


\vskip5mm

\begin{thebibliography}{99}

\bibitem{c01}
A. Linde, Phys.Rept, {\bf 333}, 575 (2000).

\bibitem{c02}
S.W.Hawking, Phys.Rev. {\bf D62}, 043501 (2000).

\bibitem{c03}
C.G.Callan, D.Friedan, E.J.Martinec, N.J.Pery, Nucl. Phys. {\bf B294},
593 (1985).

\bibitem{c04}
A.Tseytlin,
``String Solutions with Nonconstant Scalar Fields''  {\it Published in
the  proceedings   of  International  Symposium  on  Particle  Theory,
Wendisch-Rietz,     Germany,     7-11     Sep     1993     (Ahrenshoop
Symp.1993:0001-13),} hep-th/9402082

\bibitem{c05}
B.Zwiebach, Phys.Lett. {\bf B156}, 315  (1985);
E.Poisson, Class.Quant.Grav. {\bf 8}, 639 (1991);
D.Witt, Phys.Rev. {\bf D38}, 3000 (1988);
J.T.Wheeler, Nucl.Phys. {\bf B268}, 737 (1986),
             Nucl.Phys. {\bf B273}, 732 (1986);
G.W.Gibbons and K.Maeda, Nucl.Phys. {\bf B298}, 741 (1988);
D.Garfincle, G.Horowitz  and  A.Strominger,  Phys.Rev. {\bf D43}, 3140
(1991), Phys.Rev. {\bf D45}, 3888 (1992).

\bibitem{c06}
S. Mignemi and N.R. Stewart, Phys. Rev. {\bf D47}, 5259 (1993);
P. Kanti,  N.E. Mavromatos, J. Rizos,  K. Tamvakis and  E. Winstanley,
Phys. Rev. {\bf D54}, 5049 (1996);
T. Torii, H. Yajima, and K. Maeda, Phys. Rev. {\bf D55}, 739 (1997).

\bibitem{c07}
S.O. Alexeyev and M.V. Pomazanov, Phys. Rev. {\bf D55}, 2110 (1997);
S.O. Alexeyev  and M.V. Sazhin, Gen. Relativ. and  Grav. {\bf 8}, 1187
(1998);
S.O. Alexeyev, M.V. Sazhin and M.V.Pomazanov, Int. J.  Mod. Phys. {\bf
D10}, 225 (2001).

\bibitem{c071}
J.D. Bekensten, Phys. Rev. {\bf D49}, 1912 (1994).

\bibitem{c072}
A. Barrau, Astropart.Phys. {\bf 12}, 269 (2000).

\bibitem{c08}
A.G.Polnarev,  M.Yu.Khlopov,  Astronomicheskii  Zhurnal  (Astronomical
Journal) {\bf 58}, 706 (1981), in Russian;
M.Yu.Khlopov, B.A.Malomed, Ya.B.Zeldovich, MNRAS {\bf 215}, 575 (1985);
J.H.MacGibbon, Nature {\bf 329}, 308 (1987);
J.H.MacGibbon, B.Carr, Ap.J. {\bf 371}, 447 (1991);
V.I.Manko, M.A.Markov, Phys.Lett. {\bf A172}, 331 (1993).

\bibitem{c09}
P. Kanti,  N.E. Mavromatos, J. Rizos,  K. Tamvakis and  E. Winstanley,
Phys.Rev. {\bf D57}, 6255 (1998);
T.Torii, K.Maeda, Phys.Rev. {\bf D58}, 084004 (1998).

\bibitem{c10}
O.Khovanskaya, ``Dilatonic  black  hole time stability'', accepted for
publication to Grav. \& Cosmol for (2002).

\bibitem{c11}
S.Alexeyev, S.Mignemi, Class. Quant. Grav. {\bf 18}, 4165 (2001).

\bibitem{c12}
L.Randall, R.Sundrum, Phys. Rev. Lett. {\bf 83}, 4690 (1999).

\bibitem{c121}
R.R. Metsaev, A.A. Tseytlin, Phys. Lett. {\bf B 185}, 52 (1987);
M.C. Bento, O. Bertolami, Phys. Lett. {\bf B 228}, 348 (1989),
                                      {\bf B 368}, 198 (1996).

\bibitem{c122}
S.O. Alexeyev, O.S.Khovanskaya, Grav.\& Cosmol, {\bf 6}, 14 (2000).

\bibitem{c13}
P.R. Branoff, D.R. Brill,
``Instantons for black hole pair production'',
To be  published in a festschrift  for J.V. Narlikar,  Kluwer Academic
Pub., 1999, gr-qc/9811079

\bibitem{c130}
S.Alexeyev,  A.Barrau,  G.Boudoul, O.Khovanskaya,  M.Sazhin, Astronomy
Letters ({\it Pisma v Astronomichesky Zhurnal}), {\bf 28}, 428 (2002).

\bibitem{c131}
R.C.Myers, J.Z.Simon, Phys. Rev. {\bf D38}, 2434 (1988);
Gen. Rel. Grav. {\bf 21}, 761 (1989).

\bibitem{c132}
J.D.Barrow,  E.J.Copeland,  A.R.Liddle,  Phys.  Rev.  {\bf  D46},  645
(1992).

\bibitem{c139}
S. Massar \& R. Parentani, Nucl.Phys. B575, 333-356 (200)

\bibitem{c14}
M.K.Parikh, F.Wilczek, Phys. Rev. Lett. {\bf 85}, 24 (2000).

\bibitem{c14a}
K.Srinivasan, T. Padamanabhan, Phys. Rev. {\bf D60}, 24007 (1999).

\bibitem{c15}
T.Damour, R.Ruffini, Phys. Rev. {\bf D14}, 332 (1976).

\bibitem{c17}
D.N.Page, Phys.Rev. {\bf D13}, 198 (1976); {\bf D14}, 3260 (1976);
{\bf D16}, 2402 (1977).

\bibitem{c18}
W.A.Hiscock, Phys. Rev. {\bf 23}, 2823 (1981).

\bibitem{c19}
J.M.Blatt,  V.F.Weisskopf,  Theoretical Nuclear  Physics  (Wiley,  New
York, 1952), p. 520;
A.A.Starobinsky, S.M.Churilov, Zh. Eksp. Fiz. {\bf 65}, 3 (1973);
J.H.MacGibbon, B.R.Webber, Phys. Rev. {\bf D41}, 3052 (1990).

\bibitem{c20a}
S.Alexeyev, O.Khovansaya,  M.Sazhin,  Astronomy  Letters ({\it Pisma v
Astronomichesky Zhurnal}) {\bf 28}, 139 (2002).

\bibitem{c21}
C. Renault {\it  et al.}, Astronomy  \& Astrophysics, {\bf  324},  L69
(1997).

\bibitem{c22}
H.I. Kim, Phys.Rev. {\bf D62}, 063504 (2000).

\bibitem{c23}
T. Bringmann, C.  Kiefer, D. Polarski,  Phys. Rev. {\bf  D65},  024008
(2002).

\end{thebibliography}
\end{document}